# Correlation between electronic and structural orders in 1*T*-TiSe$_2$

Hiroki Ueda[1,†], Michael Porer[1,†], José R. L. Mardegan[1,a], Sergii Parchenko[1,b], Namrata Gurung[2,3], Federica Fabrizi[4], Mahesh Ramakrishnan[1], Larissa Boie[5], Martin Josef Neugebauer[5], Bulat Burganov[5], Max Burian[1], Steven Lee Johnson[5,6], Kai Rossnagel[7,8] and Urs Staub[1,*]

[1] *Swiss Light Source, Paul Scherrer Institute, 5232 Villigen-PSI, Switzerland.*

[2] *Laboratory for Multiscale Materials Experiments, Paul Scherrer Institute, 5232 Villigen-PSI, Switzerland.*

[3] *Laboratory for Mesoscopic Systems, Department of Materials, ETH Zurich, 8093 Zurich, Switzerland.*

[4] *Diamond Light Source Ltd., Diamond House, Harwell Science & Innovation Campus, Didcot, Oxfordshire OX11 0DE, United Kingdom.*

[5] *Institute for Quantum Electronics, Physics Department, ETH Zurich, 8093 Zurich, Switzerland.*

[6] *Laboratory for Non-linear Optics, Paul Scherrer Institute, 5232 Villigen-PSI, Switzerland.*

[7] *Institut für Experimentelle und Angewandte Physik, Christian-Albrechts-Universität zu Kiel, 24098 Kiel, Germany.*

[8] *Ruprecht-Haensel-Labor, Deutsches Elektronen-Synchrotron DESY, 22607 Hamburg Germany.*

The correlation between electronic and crystal structures of 1*T*-TiSe$_2$ in the charge density wave (CDW) state is studied by x-ray diffraction. Three families of reflections are used to probe atomic displacements and the orbital asymmetry in Se. Two distinct onset temperatures are found, $T_{CDW}$ and a lower $T^*$ indicative for an onset of Se out-of-plane atomic displacements. $T^*$ coincides with a DC resistivity maximum and the onset of the proposed gyrotropic electronic structure. However, no indication for chirality is found. The relation between the atomic displacements and the transport properties is discussed in terms of Ti 3*d* and Se 4*p* states that only weakly couple to the CDW order.

[a] Present address: *Deutsches Elektronen-Synchrotron DESY, 22603 Hamburg, Germany.*

[b] Present address: *Laboratory for Mesoscopic Systems, Department of Materials, ETH Zurich, 8093 Zurich, Switzerland.*

† These authors contributed equally.

* To whom correspondence should be addressed. E-mail: urs.staub@psi.ch

Electron (hole) localization are central concepts in condensed matter systems and are directly relevant for charge density wave (CDW) order, exciton formation and superconductivity. The transition-metal dichalcogenide 1$T$-TiSe$_2$ has long been studied due to its wide variety of exciting physical phenomena such as CDW formation [1,2], the possible emergence of a Bose-Einstein exciton condensate [3,4] and very recently, the observation of a gyrotropic electronic order [5]. Additionally, emergent superconductivity is found in the vicinity of the CDW phase [6–8]. The origin of the CDW, i.e., whether it is driven by an excitonic electron-hole interaction or an electron-phonon coupling [9–14], remains controversial. Consequently, the complex relationship among electronic symmetry breaking, transport properties and the deformed crystal structure attract continuous attention [1,15–21].

In the normal state, above the CDW transition temperature $T_{CDW}$ ($\approx$ 202 K), the 1$T$-TiSe$_2$ crystal structure can be well-described by space group $P\bar{3}m1$ [22]. It is a semiconductor with the Se 4$p$ valence band maximum lying at the $\Gamma$ point and the Ti 3$d$ conduction band minima at the $L$ points [see Fig. S1(a)] [23]. The CDW order is represented by three wave vectors $\mathbf{q}$ [= $\mathbf{a}$*/2+$\mathbf{c}$*/2, $-\mathbf{b}$*/2+$\mathbf{c}$*/2, and $-\mathbf{a}$*/2+$\mathbf{b}$*/2+$\mathbf{c}$*/2, where $\mathbf{a}$, $\mathbf{b}$ and $\mathbf{c}$ are hexagonal bases used in Figs. 1(a) and 1(b)], which connect the $\Gamma$ point and the $L$ points as seen in Fig. 1S(b). The formation of the triple-$q$ CDW state results in a doubling of the hexagonal lattice parameters. The 2×2×2 commensurate superlattice structure in the CDW state is well-described by space group $P\bar{3}c1$ and is shown in Figs. 1(a) and 1(b). As a result of the CDW order, band folding occurs and all of the $\Gamma$, $M$, $A$ and $L$ points in the normal state become the high-symmetric $\Gamma_{CDW}$ point in the CDW state. Moreover, the valence band maximum at the $\Gamma$ point and the conduction band minima (normal state $L$ points) are strongly hybridized at the $\Gamma_{CDW}$ point and are significantly repelled from the Fermi energy. A recent photoemission study revealed, however, that a branch of the conduction band derived from Ti 3$d_{z^2}$ states remains as in the normal state, indicating only a weak correlation to the CDW order [24]. This Ti 3$d_{z^2}$ band crosses the Fermi level and dominates transport properties in the CDW state. In contrast, the repulsion of the valence band maximum results in a valence band maximum of the CDW state that is no longer at the $\Gamma$ point but is located at the $A$ point, at which, the states are only weakly affected by the CDW order. The band structures for both the states are schematically drawn in Figs. S1 and 1(d).

DC resistivity shows an increase at $T_{CDW}$ which could be due to charge localization and gap formation, similar to other CDW systems [25–27]. However, despite the reconstruction of the band structure at $T_{CDW}$, a large and broad peak in the resistivity is seen at a temperature lower than $T_{CDW}$ [$T^* \approx 165$ K, see Fig. 2(a)]. It has been suggested that this anomalous peak originates from a crossover between two temperature regimes: from the low-temperature state with negative carriers (electrons) resulting from a small extrinsic doping to a high-temperature state with both negative and positive carriers (electrons and holes) which arise due to thermal fluctuations [21].

Besides the anomalous resistivity peak, the CDW state also features a circular photogalvanic current interpreted as the occurrence of gyrotropic electronic order [5] and the emergence of chirality [28], both of which have distinct onset temperatures below $T_{CDW}$. The chiral phase transition, as claimed to be observed by scanning tunneling microscopy (STM) and x-ray diffraction (XRD) experiments [28–30], takes place $\approx 7$ K lower than $T_{CDW}$ [28]. Such a chiral CDW order with $P1$ symmetry is manifested by chiral atomic displacements and a possible orbital order [31–35]. However, these experimental findings have been challenged by alternative interpretations [36–38]. It is, therefore, important to clarify the electronic and crystal structures and their temperature dependence.

In this Letter, we study in detail the various structural modifications below $T_{CDW}$ in a thin flake of single crystalline 1$T$-TiSe$_2$. The Bragg reflections, which correlate to the CDW order, are ascribed to specific atomic motions. The change in transport properties below $T_{CDW}$ is discussed in terms of a possible change of orbital hybridization caused by these atomic motions. Although the onset temperature of certain reflections deviate from $T_{CDW}$, no evidence of chirality [28] is found. Detailed investigations of a space-group forbidden reflection in resonant XRD combined with *ab initio* simulation are well described by a non-chiral symmetry of electronic and crystal structures in the CDW state.

Figures 1(a) and 1(b) display the crystal structure of 1$T$-TiSe$_2$ in the CDW state ($P\bar{3}c1$) [22]. Both Ti and Se atoms, each of which occupies a single crystallographic site in the normal state, become two distinct sites in the CDW state. While Ti1 and Se2 have only single in-plane ($\delta_{Ti}$) or out-of-plane ($\delta_{Se2\_out}$) displacements from the normal state, respectively, Se1 has both types of components ($\delta_{Se1\_in}$ and $\delta_{Se1\_out}$ in-plane and out-of-plane, respectively). Since there are two Se sites, the out-of-plane motions of these sites are independent and we can define their difference as $\Delta z_{Se} = \delta_{Se1\_out} - \delta_{Se2\_out}$.

A single crystal of 1$T$-TiSe$_2$ grown by the iodine vapor transport method [24] was cleaved along the (001) plane by repetitive exfoliation, and a flake ($\approx$ 2 μm thickness) was mounted on a polycrystalline diamond substrate. Experiments were performed at the beamline I16 [39] of Diamond Light Source. The sample was mounted on the cold finger of a closed cycle refrigerator attached to a Newport six-circle kappa diffractometer and cooled down well below $T_{CDW}$ and $T^*$. The photon energies of x-ray beams were tuned to 12.6 keV for non-resonant XRD and around the Se $K$ edge ($\approx$ 12.658 keV) for resonant XRD. A Si (111) analyzer was used to determine the polarization state of the scattered beam. Scattered photons were counted using a PILATUS 100K pixel detector [40] and using an avalanche photo diode during azimuthal scans at resonance. A rocking curve as well a raw two-dimensional image of (1 0 7) shown in Fig. S2 confirm the high crystalline quality of the sample. X-ray absorption spectrum (XAS) was obtained by integrating the fluorescence signal. DC resistivity was measured for a single crystal from the same batch with that used for this XRD study along the (001) plane by the four-points resistance method.

We classify the measured Bragg reflections into three families: A, B and C, all of which are space-group forbidden in the normal state. We denote reflections by using the indices in the normal state, $h$, $k$ and $l$. Family A represents ($h\ k\ l$) reflections where $l$ and at least one of $h$ or $k$ are half-integers. These reflections appear at **G** + **q**, where **G** is a reciprocal lattice vector of the normal state, and thus directly measure the displacement of the lattice from the CDW order. Family B represents (0 $k\ l$) type of reflections where $l$ is an integer while $k$ is a half-integer. Family C also represents (0 $k\ l$) type of reflections but with both $l$ and $k$ as half-integers. Note that family C are space-group forbidden even in the CDW state and are observed only at resonance (as shown later).

Figure 2 shows the temperature dependence of the integrated intensities of the three families together with the DC resistivity. The reflections (0.5 1 1.5) (family A, off-resonance) and (0 0.5 9.5) (family C, at resonance) appear at $T_{CDW}$ and their intensities grow proportional to the square root of $T_{CDW} - T$. On the other hand, the (0 0.5 8) (family B, off resonance) appears at $T^*$, which coincides with the DC resistivity maximum, and shows an approximately linear dependence with temperature (see Fig. S5 for further reflections). These differences imply that A and B are sensitive to different atomic displacements allowed in the CDW state and that the displacements that dominates B have a correlation to the transport properties. The previously reported difference in the onset temperatures ($\approx$ 7 K) of (1.5 1.5 0.5) and (2.5 1 0) [28], as well as the linear temperature dependence of (2.5 1 0) are reasonably explained by a fourth-order contribution of the *in-plane* CDW distortion occurring at $T_{CDW}$ [33]. However, the observed difference in the onset temperatures ($\approx 37$ K) in our study cannot be explained in the same way since a calculated fourth-order contribution to the intensity [black curve in Fig. 2(a)] does not fit to the experimentally obtained data for (0 0.5 8).

The (0 0.5 9.5) reflection, which is space-group forbidden in the CDW state, is observed only around the Se *K* edge as seen in Fig. 3(a). The intensity depends on the azimuthal angle, and the profile is nicely described by *ab initio* calculations performed by the FDMNES code using a model based on the crystal structure in the CDW state [22] [see Fig. 3(b)]. The polarization analysis at the specific azimuthal angle displayed in Fig. S3 is consistent with the calculated results where the reflection gives finite intensity in the $\sigma-\pi'$ channel but no signal in the $\sigma-\sigma'$ channel. On the basis of such profiles, which are well reproduced by our *ab initio* calculations, and the symmetry analyses (see Supplemental Material of [41]), the (0 0.5 9.5) space-group forbidden reflection intensity is explained by an aspheric electron distribution in the Se1 orbitals in the CDW state [42], also often called ATS (anisotropic tensor susceptibility) scattering.

The x-ray absorption at the Se *K* edge includes the excitation of an electron from the 1*s* state to the 4*p* valence state, which would be totally occupied in a fully ionic picture. Thus, the observation of the space-group forbidden reflection means that there are holes in the Se valence state in addition to the aspheric electron distribution in the Se1 4*p*, as predicted in Ref. 43 by DFT calculations. It is obvious that the change in orbital asphericity is associated with the CDW order since its onset temperature coincides with the appearance of reflections of family A. The absence of calculated intensities for artificially removed in-plane CDW distortions for the reflection further confirms this relation [41]. In addition, the anisotropic environment of the hybridized Ti 3*d* states, associated with the CDW structure, allows the formation of an excitonic state [3,4] generating holes in the Se 4*p* bands which influence the electron distribution in Se.

To identify the nature of atomic displacements dominating space-group allowed reflections A and B, numerical calculations for possible models with different types of displacements were carried out and are summarized in Table S2 (see Ref. 41 for details of the models and procedure). We find that $\delta_{Ti}$ and $\delta_{Se1\_in}$, both in-plane displacements, dominate A while $\Delta z_{Se}$, relating to an out-of-plane displacement, dominates B. Figure 4 shows the calculated intensities of the (0.5 1 1.5) and (0 0.5 8) reflections as functions of $\delta_{Ti}$ and $\Delta z_{Se}$ [those of $\delta_{Se1\_in}$, exhibiting similar behaviors with those of $\delta_{Ti}$, are shown in Fig. S6(a)]. A is largely dominated by $\delta_{Ti}$ through a quadratic contribution to the intensity (linear in the structure factor) without any contribution from $\Delta z_{Se}$. On the other hand, B is dominated by $\Delta z_{Se}$ ($\propto \Delta z_{Se}^2$) with a $10^3$ times smaller quartic contribution in $\delta_{Ti}$. The quartic contribution of $\delta_{Ti}$ to the intensity can explain the linear temperature dependence of B but cannot explain the significant intensity and the difference in onset temperature from A. Thus, it is plausible that $\Delta z_{Se}$ sets in $\approx 37$ K lower than $T_{CDW}$ and develops slower as a function of temperature while cooling than the CDW order parameters, $\delta_{Ti}$ and $\delta_{Se1\_in}$. These observations are possible when $\Delta z_{Se}$ is not the primary order parameter of the CDW phase transition. The difference in onset temperature between (1.5 1.5 0.5) and (2.5 1 0) reported earlier [28,33] are not related to our observations since (2.5 1 0) is independent of $\Delta z_{Se}$ (as $l = 0$). As shown in Ref. 33, the temperature dependence is naturally explained by the quartic behavior of the in-plane displacements. However, the anomaly in the specific heat found in that study [28] might be directly related to $\Delta z_{Se}$.

Whereas the CDW order largely affects the band structure, the maximum of the DC resistivity is not correlated to $T_{CDW}$ but $T^*$, around which the gyrotropic electronic order also appears [5]. Note that $T_{CDW}$ can vary between different samples, probably due to differences in defect density [1]. The displacements $\delta_{Ti}$ and $\delta_{Se1\_in}$ dominate the CDW distortion and $\Delta z_{Se}$ does not further break the symmetry in the CDW state. The $\Delta z_{Se}$, the magnitude of which is much smaller than $\delta_{Ti}$ and $\delta_{Se1\_in}$, creates only a tiny change in the structure of the main bands that form the hybridization gap. The onset of $\delta_{Se1\_in}$ shortens the Se1-Ti bond length compared to Se2-Ti and, hence, Se1 moves away from the adjacent Ti layer below $T^*$. This weakens the orbital hybridization between Ti $3d_{z^2}$ and Se $4p_{x,y}$ at the $A$ point, increasing the band curvatures of the specific branches [compare Figs. 1(c) and 1(d)]. Therefore, the carriers in Ti $3d_{z^2}$ and $A$-derived Se $4p_{x,y}$ acquire lighter effective masses and longer mean free paths, resulting in a decrease of DC resistivity below $T^*$. Note that the Ti $3d_{z^2}$, which crosses the Fermi energy, dominates the transport in the CDW state.

Interestingly, for both $Cu_xTiSe_2$ ($x = 0.05$) and $1T$-$TiSe_2$ at $\geq 2.82$ GPa, the anomalous peak in the DC resistivity vanishes and a superconducting state appears upon lowering the temperature [6,7]. These samples have the same space group and CDW transition as $1T$-$TiSe_2$ under ambient pressures, but the distortion $\Delta z_{Se}$ is zero in the CDW state for both the doped and pressurized samples [22]. This implies that $\Delta z_{Se}$ plays a critical role in the transport properties and might be a key parameter governing the emergence of the superconducting state.

We now comment on the possible occurrence of chirality in the CDW state. Even though the B reflections have a lower onset temperature than $T_{CDW}$, as there is no indication for a first order transition, a symmetry analysis based on the Landau theory is inconsistent with the occurrence of chirality [5]. The circular photogalvanic current, which is claimed as the evidence for a chiral electronic structure, appears at almost the same temperature as the structural modification [5]. This suggests that chirality in the electronic states is significantly coupled to the lattice, and the underlying crystal lattice should be chiral. Observation of orbital asphericity through studying the polarization dependence of space-group forbidden reflections in resonant XRD is one of the established methods to determine the presence of chirality in crystal lattices [44–46]. However, the observed azimuthal-angle dependence of the (0 0.5 9.5) reflection is very well reproduced by the *ab initio* calculations assuming the non-chiral space group $P\bar{3}c1$. Moreover, this reflection must be a symmetry allowed reflection if the space group is $P1$, as claimed in earlier studies [31–35]. The absence of an intensity for this reflection at off-resonance and in the $\sigma$–$\sigma'$ channel is shown in Figs. 3(a) and 3(b), respectively. Therefore, we can conclude that our results support neither the chirality of the Se orbitals nor the low symmetry crystal structure in the CDW state.

In summary, we examined the charge density wave (CDW) phase of $1T$-TiSe$_2$ by means of x-ray diffraction. One of the studied three families of reflections has different onset temperature compared to the others. Our detailed analysis revealed that its origin comes from the difference in relative out-of-plane motion between the two Se sites, which weakly couples to the CDW order. The different onset temperature compared to the CDW transition of the reflection is not caused by the onset of chirality. Such out-of-plane atomic displacements of Se can reduce the orbital hybridization between Ti $3d_{z^2}$ and Se $4p_{x,y}$ states, which couples only weakly to the CDW order. The consequent decrease of the effective mass of carriers results in a reduction of DC resistivity at lower temperatures. Our x-ray diffraction results provide the crucial link between the origin of the anomalous feature in transport properties and structural modifications that change the hybridization in the relevant electronic bands.

ACKNOWLEDGEMENT

We thank C. Monney and T. Jaouen for enlightening discussions. The x-ray diffraction experiments were performed at the beamline I16 (Diamond Light Source) and the X04SA beamline (Swiss Light Source). This work was supported by Swiss National Science

Foundation (Sinergia Project 'Toroidal moments' Grant No. CRSII2_147606) and its National Centers of Competence in Research, NCCR MUST and NCCR MARVEL.

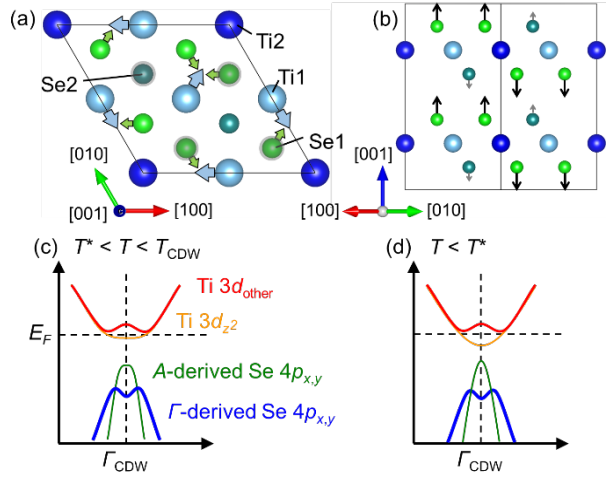

Fig. 1 Crystal structure of 1$T$-TiSe$_2$ in the CDW state viewed along (a) [001] and (b) [110] (drawn by VESTA [47]). Arrows indicate the displacement directions from the normal state. Gray circles in (a) show Se in the lower layer with respect to the adjacent Ti layer. Schematic band structures of 1$T$-TiSe$_2$ at (c) $T^* < T < T_{CDW}$ and (d) $T < T^*$. (d) is similar to a sketch shown in Ref. 24. $T_{CDW}$ and $T^*$ denote the CDW transition temperature and the onset temperature of reflections of family B (see text), respectively.

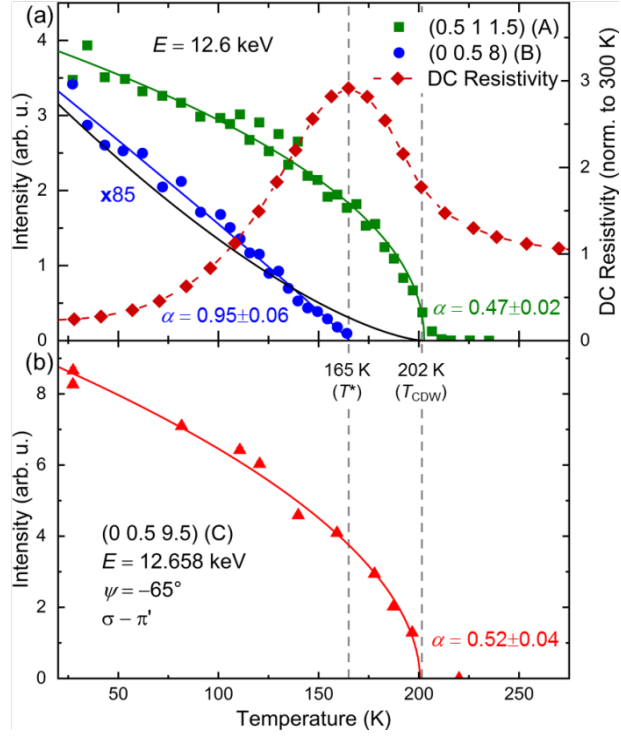

Fig. 2 Integrated intensities of reflections belonging to families A (0.5 1 1.5), B (0 0.5 8) and C (0 0.5 9.5), and DC resistivity as functions of temperature. XRD data shown in (a) [(b)] were acquired at 12.6 keV [the Se $K$ edge]. The DC resistivity is normalized by the data at 300 K. The green, red and blue solid lines represent power-law fits [$\propto (T_C - T)^\alpha$], where $T_C$ and $\alpha$ are the critical temperature and critical exponent, respectively. The black line represents the quartic term with an onset at $T_{CDW}$.

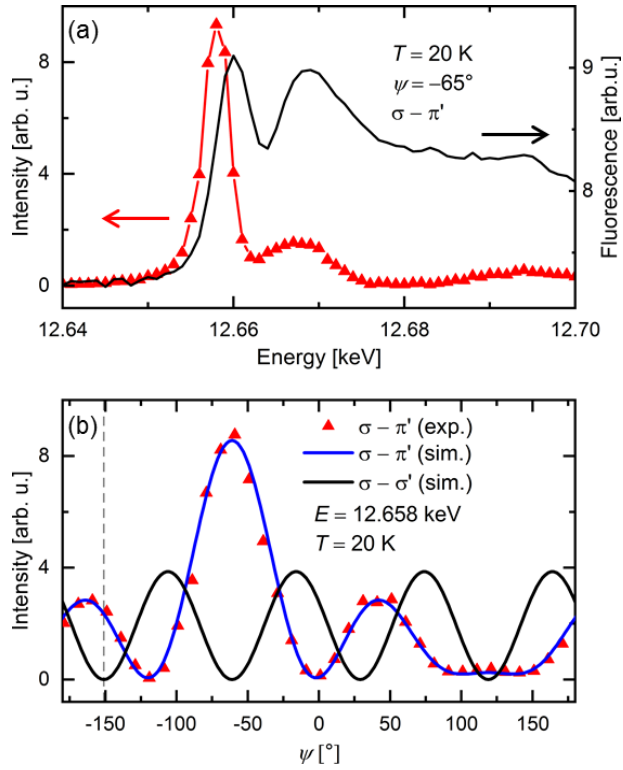

Fig. 3 (a) Photon-energy and (b) azimuthal-angle dependence of the (0 0.5 9.5). XAS (black line) is also shown in (a). Blue and black curves in (b) show the result of simulation using the FDMNES code [48]. Gray dotted line in (b) indicates the azimuthal angle from where the data for Fig. S3(a) was taken.

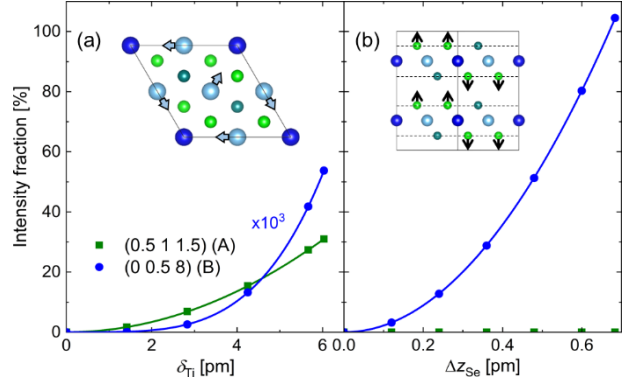

Fig. 4 Numerically calculated intensities of (0.5 1 1.5) (green) and (0 0.5 8) (blue) reflections as functions of (a) $\delta_{Ti}$ and (b) $\Delta z_{Se}$. The data are normalized by the intensity calculated for the reported structure in the CDW state [22]. The respective displacements are schematically displayed as insets. The green line in (a) and blue line in (b) represent a quadratic and the blue line in (a) a quartic function.

Supplemental Material for

# "Correlation between electronic and structural orders in 1$T$-TiSe$_2$"

by

Hiroki Ueda[1,†], Michael Porer[1,†], José R. L. Mardegan[1,a], Sergii Parchenko[1,b], Namrata Gurung[2,3], Federica Fabrizi[4], Mahesh Ramakrishnan[1], Larissa Boie[5], Martin Josef Neugebauer[5], Bulat Burganov[5], Max Burian[1], Steven Lee Johnson[5,6], Kai Rossnagel[7,8] and Urs Staub[1,*]

[1] *Swiss Light Source, Paul Scherrer Institute, 5232 Villigen-PSI, Switzerland.*

[2] *Laboratory for Multiscale Materials Experiments, Paul Scherrer Institute, 5232 Villigen-PSI, Switzerland.*

[3] *Laboratory for Mesoscopic Systems, Department of Materials, ETH Zurich, 8093 Zurich, Switzerland.*

[4] *Diamond Light Source Ltd., Diamond House, Harwell Science & Innovation Campus, Didcot, Oxfordshire OX11 0DE, United Kingdom.*

[5] *Institute for Quantum Electronics, Physics Department, ETH Zurich, 8093 Zurich, Switzerland.*

[6] *Laboratory for Non-linear Optics, Paul Scherrer Institute, 5232 Villigen-PSI, Switzerland.*

[7] *Institut für Experimentelle und Angewandte Physik, Christian-Albrechts-Universität zu Kiel, 24098 Kiel, Germany.*

[8] *Ruprecht-Haensel-Labor, Deutsches Elektronen-Synchrotron DESY, 22607 Hamburg Germany.*

[†] These authors contributed equally.

[*] To whom correspondence should be addressed. E-mail: urs.staub@psi.ch

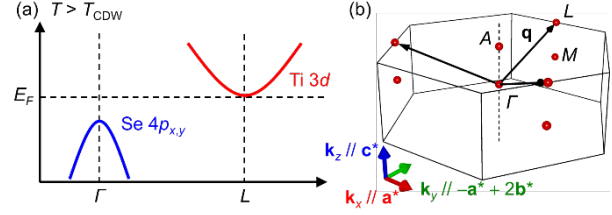

Fig. S1  (a) Schematic band structure of 1$T$-TiSe$_2$ at (d) $T > T_{CDW}$. (b) Brillouin zone of 1$T$-TiSe$_2$ in the normal state. (a) is a similar sketch shown in Ref. 1. $T_{CDW}$ denotes the CDW transition temperature.

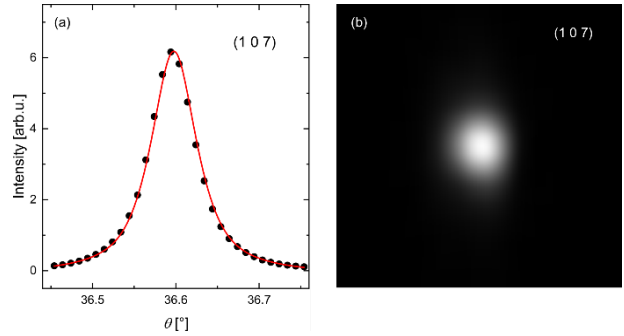

Fig. S2  (a) A rocking curve and (b) raw two-dimensional image of the (1 0 7) reflection taken at 12.6 keV and room temperature. The red curve in (a) is a Voigt function, and error bars are smaller than the marker.

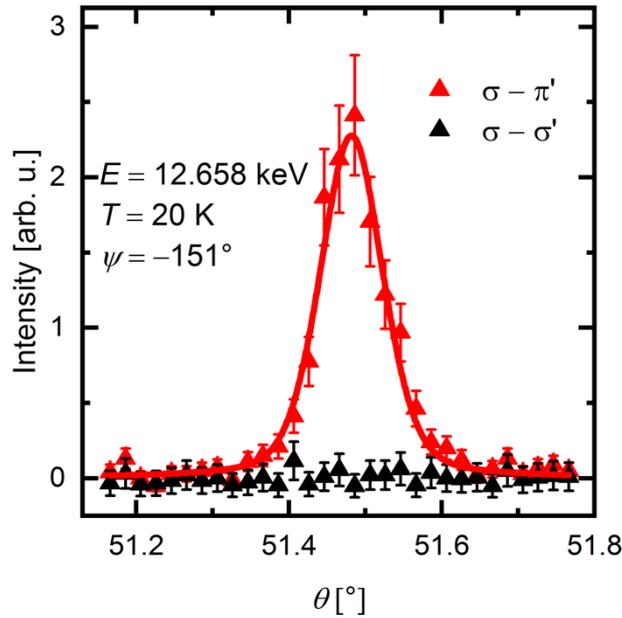

Fig. S3  Polarization analysis of the (0 0.5 9.5) reflection. The red curve is a Voigt function. These data were taken at the azimuthal angle indicated by a gray dotted line in Fig. 3(b).

I. Symmetry analyses and calculations for the (0 0.5 9.5) reflection (family C)

In order to clarify the origin of the space-group forbidden (0 0.5 9.5) reflection (family C), we show our symmetry analysis and results of *ab initio* and DFT calculations. In general, due to intra-atomic excitation of an electron between specific orbitals in the excited atoms, resonant x-ray diffraction is sensitive to the anisotropic electron distribution or orbital asphericity. This anisotropic scattering is described by an x-ray susceptibility tensor $\hat{f}$, which is defined by the global and local symmetries of the atoms [2], and leads to the emergence of a space-group forbidden reflection. We perform the symmetry analyses for Se1 and Se2, respectively, as different crystallographic sites give space-group allowed reflections while our discussion is dedicated to reveal the origin of the space-group forbidden reflection.

At first, we focus on the high-symmetry Se2 site, locating at the Wyckoff position 4$d$ with the three-fold rotational symmetry $C_3$ along [001]. With use of the Cartesian coordinate system where $x$ is along [100], $y$ is along [120] and $z$ is along [001] [see Fig. S4(e)], $\hat{f}$ is denoted as

$$\hat{f} = \begin{pmatrix} f_{xx} & f_{xy} & f_{xz} \\ f_{xy} & f_{yy} & f_{yz} \\ f_{xz} & f_{yz} & f_{zz} \end{pmatrix}. \quad (1)$$

The local $C_3$ symmetry requires the relation, $\hat{f} = C_3 \hat{f} C_3^{-1}$, so that one finds $f_{xx} = f_{yy}$ and $f_{xy} = f_{yz} = f_{xz} = 0$, meaning that the symmetry-adopted $\hat{f}$ possesses only diagonal components

$$\hat{f} = \begin{pmatrix} f_{xx} & 0 & 0 \\ 0 & f_{yy} & 0 \\ 0 & 0 & f_{zz} \end{pmatrix}. \quad (2)$$

Each position of Se2 in the unit cell of the CDW state is connected by the identical operation 1 [position 1: $\mathbf{r}_1 = (1/3, 2/3, z)$], the inversion operation $\bar{1}$ [position 2: $\mathbf{r}_2 = (2/3, 1/3, -z + 1/2)$], the $c$ glide operation $c_g$ [position 3: $\mathbf{r}_3 = (2/3, 1/3, -z)$] or the combined operation of them $\bar{1} \otimes c_g$ [position 4: $\mathbf{r}_4 = (1/3, 2/3, z + 1/2)$], and algebra shows all of the Se2 have same $\hat{f}$ [one can check by calculating $\bar{1}\hat{f}\bar{1}^{-1}$, $c_g \hat{f} c_g^{-1}$ and $(\bar{1} \otimes c_g)\hat{f}(\bar{1} \otimes c_g)^{-1}$]. The form factor $\hat{F}$ at the scattering vector $\mathbf{Q} = (0,1,19)$ [equivalent to (0,0.5,9.5) with the notation in the normal state] is obtained by summing up $\hat{f}$ at all positions in a crystal with a phase factor,

$$\hat{F} = \sum_j \hat{f}_j \exp(i\mathbf{Q} \cdot \mathbf{r}_j)$$

$$= N\hat{f}[\exp(i\mathbf{Q} \cdot \mathbf{r}_1) + \exp(i\mathbf{Q} \cdot \mathbf{r}_2) + \exp(i\mathbf{Q} \cdot \mathbf{r}_3) + \exp(i\mathbf{Q} \cdot \mathbf{r}_4)] = \mathbf{0}. \quad (3)$$

Here $N$ is the number of unit cells in the crystal. Thereby, Se2 is found to be independent of the (0 0.5 9.5) reflection. This is cross-checked from the result of *ab initio* calculation done by using the FDMNES code [3] and simply removing Se1 from the reported crystal structure in the CDW state [4] [see Fig. S4(a)].

The low-symmetry Se1 site locates at the Wyckoff position 12g, which is the general position of space group $P\bar{3}c1$ and has the local symmetry 1. Thus, all the components in $\hat{f}$ are symmetry adapted. Positions, symmetry relations, phase factors in $\hat{F}$ and $\hat{f}$ are given for all Se1 in the unit cell in Table S1. From the parameters shown in the table, one finds that $\hat{F}$ is denoted as

$$\hat{F} = \sum_j \hat{f}_j \exp(i\mathbf{Q} \cdot \mathbf{r}_j)$$

$$= N\big(\hat{f}_1 e^{i\varphi_1} + \hat{f}_2 e^{i\varphi_2} + \hat{f}_3 e^{i\varphi_3} - \hat{f}_4 e^{-i\varphi_3} - \hat{f}_5 e^{-i\varphi_1} - \hat{f}_6 e^{-i\varphi_2} + \hat{f}_1 e^{-i\varphi_1} + \hat{f}_2 e^{-i\varphi_2} + \hat{f}_3 e^{-i\varphi_3} - \hat{f}_4 e^{i\varphi_3} - \hat{f}_5 e^{i\varphi_1} - \hat{f}_6 e^{i\varphi_2}\big)$$

$$= 2N\big[(\hat{f}_1 - \hat{f}_5)\cos\varphi_1 + 2(\hat{f}_2 - \hat{f}_6)\cos\varphi_2 + 2(\hat{f}_3 - \hat{f}_4)\cos\varphi_3\big], \quad (4)$$

where $\varphi_1 = 2\pi(y + 19z)$, $\varphi_2 = 2\pi(x - y + 19z)$ and $\varphi_3 = 2\pi(-x + 19z)$. Algebra gives respective components appearing in Eq. (4) as

$$\hat{f}_1 - \hat{f}_5 = 2\begin{pmatrix} 0 & f_{xy} & f_{xz} \\ f_{xy} & 0 & 0 \\ f_{xz} & 0 & 0 \end{pmatrix}, \quad (5)$$

$$\hat{f}_2 - \hat{f}_6 = 2\begin{pmatrix} 0 & -\frac{\sqrt{3}}{4}f_{xx} - \frac{1}{2}f_{xy} + \frac{\sqrt{3}}{4}f_{yy} & -\frac{1}{2}f_{xz} - \frac{\sqrt{3}}{2}f_{yz} \\ -\frac{\sqrt{3}}{4}f_{xx} - \frac{1}{2}f_{xy} + \frac{\sqrt{3}}{4}f_{yy} & 0 & 0 \\ -\frac{1}{2}f_{xz} - \frac{\sqrt{3}}{2}f_{yz} & 0 & 0 \end{pmatrix} \text{ and} \quad (6)$$

$$\hat{f}_3 - \hat{f}_4 = 2\begin{pmatrix} 0 & \frac{\sqrt{3}}{4}f_{xx} - \frac{1}{2}f_{xy} - \frac{\sqrt{3}}{4}f_{yy} & -\frac{1}{2}f_{xz} + \frac{\sqrt{3}}{2}f_{yz} \\ \frac{\sqrt{3}}{4}f_{xx} - \frac{1}{2}f_{xy} - \frac{\sqrt{3}}{4}f_{yy} & 0 & 0 \\ -\frac{1}{2}f_{xz} + \frac{\sqrt{3}}{2}f_{yz} & 0 & 0 \end{pmatrix}. \quad (7)$$

From Eqs. (4)–(7), $\hat{F}$ is found to possess only the $xy$ ($F_{xy}$) and $xz$ ($F_{xz}$) components and is represented as

$$\hat{F} = \begin{pmatrix} 0 & F_{xy} & F_{xz} \\ F_{xy} & 0 & 0 \\ F_{xz} & 0 & 0 \end{pmatrix}. \quad (8)$$

Hereby we use the Cartesian coordinate $\xi\eta\zeta$ defined by the diffraction geometry; $\eta$ is normal to the scattering plane, $\zeta$ is along **Q** and $\xi$ is normal to both the directions [see Fig. S4(e)]. Polarization vectors of incoming and outgoing x-ray beams are denoted as $\boldsymbol{\sigma} = \boldsymbol{\sigma}' = (0,1,0)$, $\boldsymbol{\pi} = (\cos\theta, 0, -\sin\theta)$ and $\boldsymbol{\pi}' = (\cos\theta, 0, \sin\theta)$, where $\theta$ is the Bragg angle. Since **Q** = (0,0.5,9.5) is tilted from [001] (// $z$) to [120] (// $y$) by a certain angle $\chi$, $\hat{F}$ is transformed as

$$\widehat{F_{\xi\eta\zeta}} = \begin{pmatrix} 1 & 0 & 0 \\ 0 & \cos\chi & -\sin\chi \\ 0 & \sin\chi & \cos\chi \end{pmatrix} \begin{pmatrix} 0 & F_{xy} & F_{xz} \\ F_{xy} & 0 & 0 \\ F_{xz} & 0 & 0 \end{pmatrix} \begin{pmatrix} 1 & 0 & 0 \\ 0 & \cos\chi & \sin\chi \\ 0 & -\sin\chi & \cos\chi \end{pmatrix}$$

$$= \begin{pmatrix} 0 & F_{xy}\cos\chi - F_{xz}\sin\chi & F_{xy}\sin\chi + F_{xz}\cos\chi \\ F_{xy}\cos\chi - F_{xz}\sin\chi & 0 & 0 \\ F_{xy}\sin\chi + F_{xz}\cos\chi & 0 & 0 \end{pmatrix}$$

$$= \begin{pmatrix} 0 & F_{\xi\eta} & F_{\xi\zeta} \\ F_{\xi\eta} & 0 & 0 \\ F_{\xi\zeta} & 0 & 0 \end{pmatrix}, \quad (9)$$

where we set [100] along the scattering plane. The azimuthal angle $\psi$ is defined along **Q**, and $\psi$ dependence of $\widehat{F_{\xi\eta\zeta}}$ is given as

$$\widehat{F_{\xi\eta\zeta}}(\psi) = \begin{pmatrix} \cos\psi & -\sin\psi & 0 \\ \sin\psi & \cos\psi & 0 \\ 0 & 0 & 1 \end{pmatrix} \begin{pmatrix} 0 & F_{\xi\eta} & F_{\xi\zeta} \\ F_{\xi\eta} & 0 & 0 \\ F_{\xi\zeta} & 0 & 0 \end{pmatrix} \begin{pmatrix} \cos\psi & \sin\psi & 0 \\ -\sin\psi & \cos\psi & 0 \\ 0 & 0 & 1 \end{pmatrix}$$

$$= \begin{pmatrix} -F_{\xi\eta}\sin 2\psi & F_{\xi\eta}\cos 2\psi & F_{\xi\zeta}\cos\psi \\ F_{\xi\eta}\cos 2\psi & F_{\xi\eta}\sin 2\psi & F_{\xi\zeta}\sin\psi \\ F_{\xi\zeta}\cos\psi & F_{\xi\zeta}\sin\psi & 0 \end{pmatrix} \quad (10)$$

and is decomposed at the respective channels,

$$\widehat{F_{\xi\eta\zeta}^{\sigma-\sigma'}}(\psi) = \begin{pmatrix} 0 & 1 & 0 \end{pmatrix} \begin{pmatrix} -F_{\xi\eta}\sin 2\psi & F_{\xi\eta}\cos 2\psi & F_{\xi\zeta}\cos\psi \\ F_{\xi\eta}\cos 2\psi & F_{\xi\eta}\sin 2\psi & F_{\xi\zeta}\sin\psi \\ F_{\xi\zeta}\cos\psi & F_{\xi\zeta}\sin\psi & 0 \end{pmatrix} \begin{pmatrix} 0 \\ 1 \\ 0 \end{pmatrix}$$

$$= F_{\xi\eta}\sin 2\psi \text{ and} \quad (11)$$

$$\widehat{F_{\xi\eta\zeta}^{\sigma-\pi'}}(\psi) = \begin{pmatrix} \cos\theta & 0 & \sin\theta \end{pmatrix} \begin{pmatrix} -F_{\xi\eta}\sin 2\psi & F_{\xi\eta}\cos 2\psi & F_{\xi\zeta}\cos\psi \\ F_{\xi\eta}\cos 2\psi & F_{\xi\eta}\sin 2\psi & F_{\xi\zeta}\sin\psi \\ F_{\xi\zeta}\cos\psi & F_{\xi\zeta}\sin\psi & 0 \end{pmatrix} \begin{pmatrix} \cos\theta \\ 0 \\ \sin\theta \end{pmatrix}$$

$$= F_{\xi\eta}\cos\theta\cos 2\psi + F_{\xi\zeta}\sin\theta\sin\psi. \quad (12)$$

From Eqs. (11) and (12), one finds the intensity at the channels as

$$I^{\sigma-\sigma'}(\psi) = \frac{|F_{\xi\eta}|^2}{2}(1 - \cos 4\psi) \text{ and} \quad (13)$$

$$I^{\sigma-\pi'}(\psi) = \frac{1}{2}\left(|F_{\xi\eta}|^2\cos^2\theta - |F_{\xi\varsigma}|^2\sin^2\theta\right)\cos 4\psi + \frac{1}{4}\left(F_{\xi\eta}F_{\xi\varsigma}^* + F_{\xi\eta}^*F_{\xi\varsigma}\right)\sin 2\theta \sin 3\psi - \frac{1}{4}\left(F_{\xi\eta}F_{\xi\varsigma}^* + F_{\xi\eta}^*F_{\xi\varsigma}\right)\sin 2\theta \sin \psi + \frac{1}{2}\left(|F_{\xi\eta}|^2\cos^2\theta + |F_{\xi\varsigma}|^2\sin^2\theta\right). \quad (14)$$

The experimental data as well the results of *ab initio* calculations shown in Fig. 3(b) are reproduced by Eqs. (13) and (14) [see Fig. S4(c)]. Furthermore, *ab initio* calculation with removing Se2 from the crystal structure in the CDW state [4] shown in Fig. S4(a) cross-checks the results of our symmetry analysis. Therefore, Se1 is found to be responsible to the space-group forbidden (0 0.5 9.5) reflection. We remark that the (0 0.5 9.5) reflection is space-group forbidden and, thereby, the scattering for the reflection is not caused by a charge but an aspheric orbital (or a quadrupole moment), whose orientation is modulated in the unit cell.

Table S1 Atomic position, symmetry relation, the phase factor in the form factor at the scattering vector $\mathbf{Q} = (0,1,19)$ [$(0,0.5,9.5)$ in the normal state] and x-ray susceptibility tensor of Se1 in the CDW state.

| Number | Position | Symmetry connection | Phase factor | $\hat{f}$ |
|---|---|---|---|---|
| 1 | $\mathbf{r}_1 = (x,y,z)$ | 1 | $e^{i\varphi_1}$ | $\hat{f}_1$ |
| 2 | $\mathbf{r}_2 = (-y, x-y, z)$ | $C_3$ | $e^{i\varphi_2}$ | $\hat{f}_2 = C_3 \hat{f}_1 C_3^{-1}$ |
| 3 | $\mathbf{r}_3 = (-x+y, -x, z)$ | $C_3^2$ | $e^{i\varphi_3}$ | $\hat{f}_3 = C_3^2 \hat{f}_1 C_3^{2\,-1}$ |
| 4 | $\mathbf{r}_4 = (y, x, -z+1/2)$ | $2_{[110]}$ | $-e^{-i\varphi_3}$ | $\hat{f}_4 = 2_{[110]} \hat{f}_1 2_{[110]}^{-1}$ |
| 5 | $\mathbf{r}_5 = (x-y, -y, -z+1/2)$ | $2_{[100]}$ | $-e^{-i\varphi_1}$ | $\hat{f}_5 = 2_{[100]} \hat{f}_1 2_{[100]}^{-1}$ |
| 6 | $\mathbf{r}_6 = (-x, -x+y, -z+1/2)$ | $2_{[010]}$ | $-e^{-i\varphi_2}$ | $\hat{f}_6 = 2_{[010]} \hat{f}_1 2_{[010]}^{-1}$ |
| 7 | $\mathbf{r}_7 = (-x, -y, -z)$ | $1 \otimes \bar{1}$ | $e^{-i\varphi_1}$ | $\hat{f}_1$ |
| 8 | $\mathbf{r}_8 = (y, -x+y, -z)$ | $C_3 \otimes \bar{1}$ | $e^{-i\varphi_2}$ | $\hat{f}_2$ |
| 9 | $\mathbf{r}_9 = (x-y, x, z)$ | $C_3^2 \otimes \bar{1}$ | $e^{-i\varphi_3}$ | $\hat{f}_3$ |
| 10 | $\mathbf{r}_{10} = (-y, -x, z + 1/2)$ | $2_{[110]} \otimes \bar{1}$ | $-e^{i\varphi_3}$ | $\hat{f}_4$ |
| 11 | $\mathbf{r}_{11} = (-x + y, y, z + 1/2)$ | $2_{[100]} \otimes \bar{1}$ | $-e^{i\varphi_1}$ | $\hat{f}_5$ |
| 12 | $\mathbf{r}_{12} = (x, x - y, z + 1/2)$ | $2_{[010]} \otimes \bar{1}$ | $-e^{i\varphi_2}$ | $\hat{f}_6$ |

By removing all the in-plane displacements appearing at the CDW order, i.e., for both Ti1 and Se1, we perform the *ab initio* calculations for the (0 0.5 9.5) reflection but no significant intensity is obtained at both the channels [see Fig. S4(b)]. The results mean that the in-plane displacements cause an aspheric and spatially modulated orbital state at Se1, which is responsible for the (0 0.5 9.5) reflection. This is supported by the onset temperature of reflections of family C that is the same as that of A (at $T_{CDW}$) as shown in Fig. 2.

Visualization of the electronic density distribution of states involved to the CDW transition yields an intuitive picture of the asphericity of the orbitals [5]. Figure S4(d) shows an isosurface of the electronic density formed by occupied states in an energy window between the Fermi energy $E_F$ and $E_F - 54$ meV. Hybridization of (occupied) Ti 3$d$ and Se 4$p$ states occurs mostly within this energy window and appears as distinctive "connecting" density between neighboring Se1 and Ti1 sites. The pattern of the density asymmetry at the Se 1 site corresponds to the asymmetry pattern of which the resonant x-ray diffraction experiment probes the Se 4$p$ quadrupole (selected by the dipole transition) that gives rise to the reflections of family C. The DFT calculation reproduces the results of [5] and is performed using the *elk* code [6] [Version 4.3.6 using generalized gradient approximation (PBEsol functional) and standard quality settings, except setting an 9×9×5 $k$-point sampling of the $P\bar{3}c1$ Brillouin zone and the product of the average muffin tin radius and the maximum reciprocal lattice vector to 8.5].

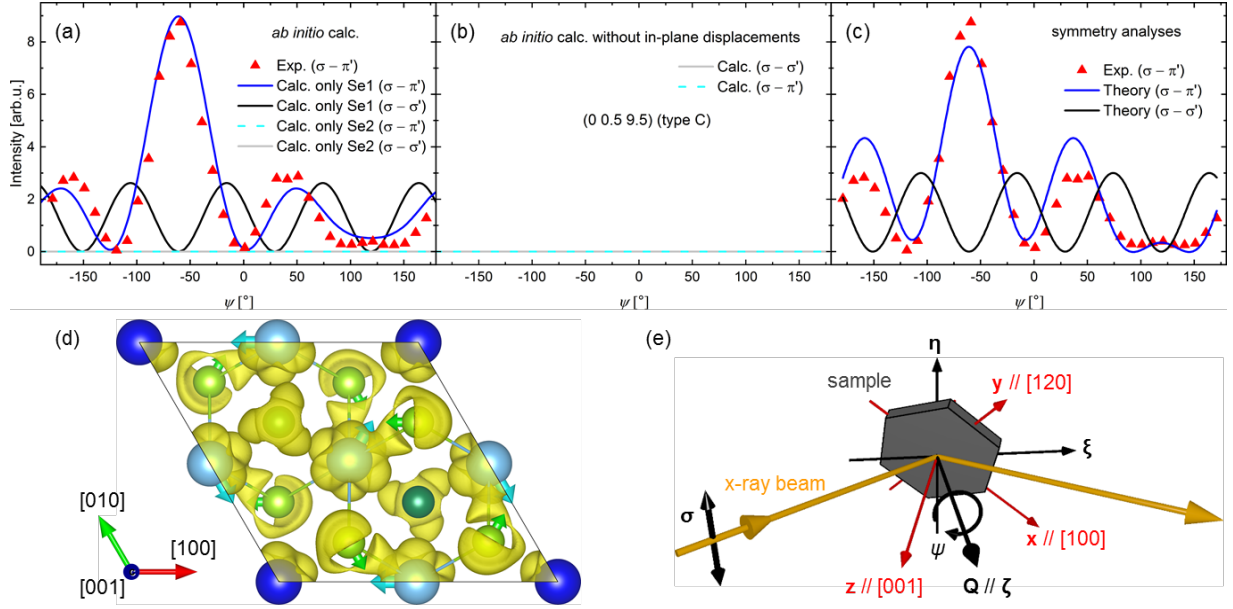

Fig. S4 (a),(b) Results of several *ab initio* calculations based on the FDMNES code for the (0 0.5 9.5) reflection with an artificial change in the crystal structure of the CDW state: (a) removing either Se1 or Se2 from the crystal structure in the CDW state and (b) removing all the in-plane displacements emerging at the CDW order, i.e., for Ti1 and Se1. (c) Azimuthal angle dependence of the (0 0.5 9.5) reflection from the results of the symmetry analyses. The experimental data shown in (a) and (c) is reproduced from Fig. 3(b). (d) Isosurface (yellow) of the electronic density of states for states lying within an energy slice between the Fermi energy $E_F$ and $E_F - 54$ meV (DFT calculation) superimposed to the crystal structure shown in Fig. 1(a). (e) Diffraction geometry and Cartesian coordinate systems used in the symmetry analyses.

II. Characterization of further reflections

In addition to the (0.5 1 1.5) and (0 0.5 8) reflections, which are shown in the main text, we characterized further reflections. Another flake of a single crystal of $1T$-TiSe$_2$ was measured at the X04SA beamline of Swiss Light Source [7]. The sample was cooled down well below $T_{CDW}$ and $T^*$ by using a N$_2$ cryoblower. The photon energy of x-ray beams was tuned to 9.2 keV, and signals were detected by a PILATUS II photon-counting pixel detector [8].

Figure S5 shows the temperature dependence of (2.5 0.5 1.5) (A), (0 0.5 2) (B) and (1.5 0 2) (B) reflections. While the (2.5 0.5 1.5) reflection onsets at ≈ 191 K {= $T_{CDW}$, note that $T_{CDW}$ can vary between different samples (defect dependent) [9]} and develops as an almost square root upon temperature for cooling, the (0 0.5 2) and (1.5 0 2) reflections have two temperature ranges; one is in the range of $T^* < T < T_{CDW}$ and the other one is $T < T^*$, where $T^*$ ≈ 167 K. Note that $T^*$ is consistent with the onset temperature of the relative out-of-plane displacements of two Se shown in the main text (165 K). In both the ranges the intensities show linear temperature dependence for cooling but have different slope with a considerable increase below $T^*$. The red and blue solid curves are a fit with the fourth-order contribution, which onsets at $T_{CDW}$, and the second-order contribution, which onsets at $T^*$. The well-reproducible results indicate the presence of the two contributions for B: the quartic contribution from the in-plane CDW distortion and the quadratic contribution from the relative out-of-plane displacements between two Se.

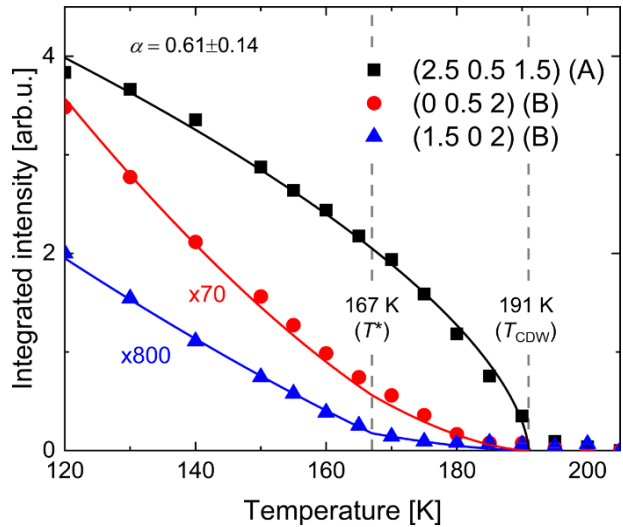

Fig. S5  Integrated intensities of three reflections, (2.5 0.5 1.5), (0 0.5 2) and (1.5 0 2), as functions of temperature. The black solid line represents a power-law fit [$\propto (T_C - T)^\alpha$], where $T_C$ and $\alpha$ are the critical temperature and critical exponent, respectively. Red and blue solid curves represent the sum of the quartic term with an onset at $T_{CDW}$ and the quadratic term with an onset at $T^*$.

III.     Numerical calculations of diffraction intensity

The calculations of diffraction intensities for the space-group allowed reflections, A and B, were carried out with several model structures having atomic displacements from the normal state. The first and second ones have only the in-plane displacements of Ti1 ($\delta_{Ti}$) and Se1 ($\delta_{Se1\_in}$) along the hexagonal axes, i.e., [100], [010], and [$\bar{1}\bar{1}0$], respectively. Whereas $\delta_{Ti}$ is along the hexagonal axes, $\delta_{Se1\_in}$ is off ($\approx 0.2°$) the hexagonal axes. Thus, we examined the third model structure having only the off-axis displacements of Se1 $\delta_{Se1\_in\_off}$. The fourth one has different $z$ coordinates between Se1 and Se2 ($\Delta z_{Se} \neq 0$). As the two Se sites have general value of $z$ in the normal state with symmetrical restriction to possess same $z$ between them, there is no contribution to any family of reflection by changing $z$ of the two Se sites as long as $\Delta z_{Se} = 0$. We individually changed the atomic positions from the crystal structure in the normal state [4] and calculated diffraction intensities of the reflections measured in our XRD study and Ref. 10. Here the x-ray energy for the calculation is identical to that used in our experiments or Ref. 10. The results are tabulated in Table S2 and shown in Figs. 4 [for (0.5 1 1.5) and (0 0.5 8) with the first and fourth models] and S6 (the other cases).

At first, it is clear from Fig. S6(e) that the in-plane off-axes displacements show tiny contribution for any reflections than the others. The reflections of family A are dominated by the in-plane displacements of Ti and Se along the hexagonal axes ($\propto \delta_{Ti}^2, \delta_{Se1\_in}^2$) without any contribution from the relative out-of-plane displacements of two Se as seen in Table S2. On the other hand, except the (1.5 0 2) reflection, which has a very weak diffraction intensity, the reflections of family B are basically dominated by the out-of-plane displacements ($\propto \Delta z_{Se}^2$) with $10^3$ or $10^4$ times smaller quartic contribution from the in-plane displacements ($\propto \delta_{Ti}^4$, $\delta_{Se1\_in}^4$). The (2.5 1 0) reflection, which has a different onset temperature compared to (1.5 1.5 0.5) on first inspection and was discussed as the indication of a chiral phase transition [10], is not classified into any of the three families and is dominated by the in-plane displacements with a quartic contribution ($\propto \delta_{Ti}^4$, $\delta_{Se1\_in}^4$) as reported in Ref. 11.

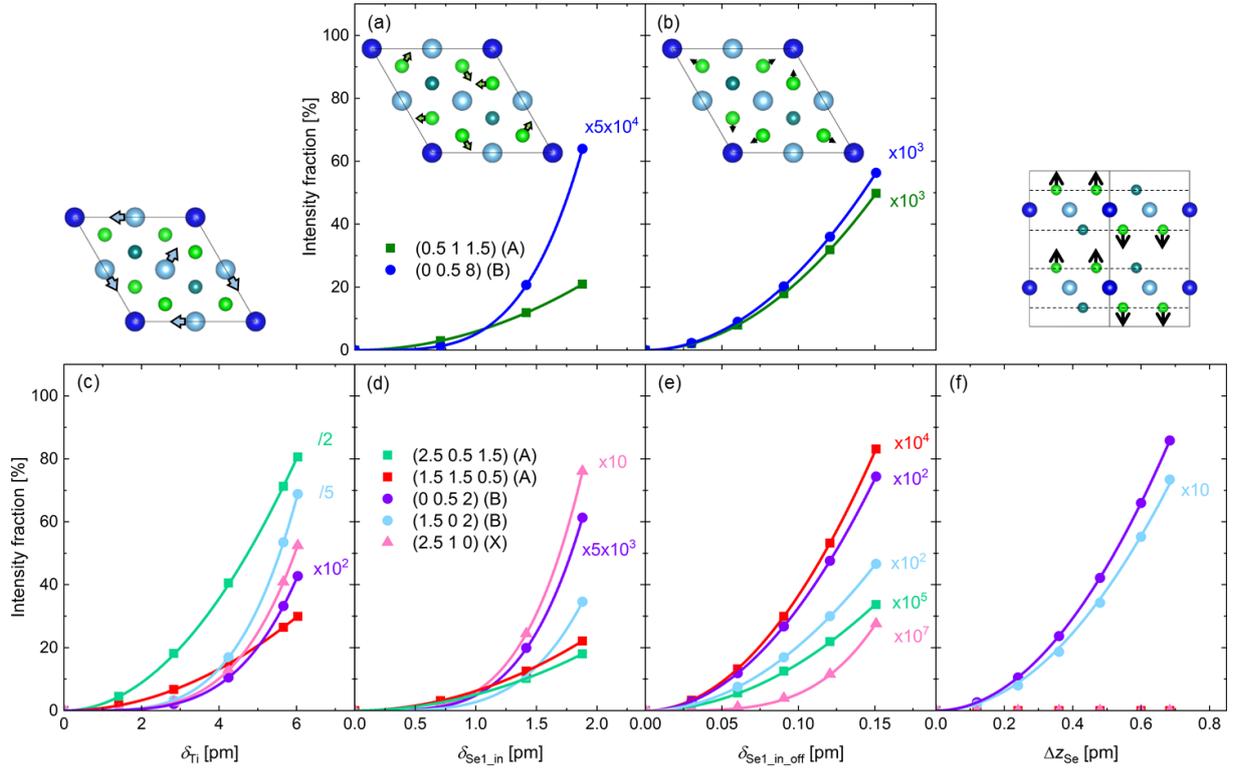

Fig. S6 Calculated diffraction intensities of space-group allowed reflections, A and B, as functions of the in-plane Se displacements along [(a) and (d)] and normal to [(b) and (e)] the in-plane hexagonal axes, (c) the in-plane Ti displacements and (f) the relative out-of-plane Se displacements. The displacements are schematically drawn. Square, circle and triangle symbols represent the reflections of family A, family B and (2.5 1 0), respectively, and solid curves are a quadratic or quartic function.

Table S2 Calculated diffraction intensities from the model structures. $F$ denotes an intensity fraction, dividing the intensities by those calculated from the crystal structure in the CDW state [4]. The family C reflection is space-group forbidden in the CDW state.

| Type | Reflection | Model 1 | | Model 2 | | Model 3 | | Model 4 | |
|---|---|---|---|---|---|---|---|---|---|
| | | $I$ [arb.u.] | $F$ [%] | $I$ [arb.u.] | $F$ [%] | $I$ [arb.u.] | $F$ [%] | $I$ [arb.u.] | $F$ [%] |
| A | (0.5 1 1.5) | 40.7 | 31.0 | 27.5 | 20.1 | $6.5 \times 10^{-2}$ | $5.0 \times 10^{-2}$ | N.D. | N.D. |
| A | (2.5 0.5 1.5) | 160.3 | 161.2 | 17.9 | 18.0 | $3.4 \times 10^{-4}$ | $3.4 \times 10^{-4}$ | N.D. | N.D. |
| A | (1.5 1.5 0.5) | 172.1 | 29.9 | 127.2 | 22.1 | $4.5 \times 10^{-2}$ | $8.3 \times 10^{-3}$ | N.D. | N.D. |
| B | (0 0.5 8) | $2.0 \times 10^{-3}$ | $5.4 \times 10^{-2}$ | $4.8 \times 10^{-5}$ | $1.3 \times 10^{-3}$ | $2.1 \times 10^{-3}$ | $5.6 \times 10^{-2}$ | 3.9 | 104.5 |
| B | (0 0.5 2) | $9.4 \times 10^{-3}$ | $4.3 \times 10^{-1}$ | $2.7 \times 10^{-4}$ | $1.2 \times 10^{-1}$ | $1.6 \times 10^{-2}$ | $7.4 \times 10^{-1}$ | 1.9 | 85.8 |
| B | (1.5 0 2) | $5.0 \times 10^{-1}$ | 344.0 | $5.0 \times 10^{-2}$ | 34.6 | $6.7 \times 10^{-4}$ | $4.7 \times 10^{-1}$ | $1.1 \times 10^{-2}$ | 7.3 |
| C | (0 0.5 9.5) | N.D. | N.D. | N.D. | N.D. | N.D. | N.D. | N.D. | N.D. |
| X | (2.5 1 0) | 3.2 | 52.4 | $4.6 \times 10^{-1}$ | 7.6 | $1.7 \times 10^{-7}$ | $2.8 \times 10^{-6}$ | N.D. | N.D. |